\documentclass[12pt,preprint]{aastex}   

\shorttitle{Latitudinal variation of the solar intensity}

\begin{document}

\title{Latitudinal variation of the solar photospheric intensity}

\author{Mark P. Rast}
\affil{Laboratory for Atmospheric and Space Physics, Department of Astrophysical and
Planetary Sciences, University of Colorado, Boulder, CO 80309 USA}
\affil{$\rm{and}$}
\affil{High Altitude Observatory, National Center for Atmospheric
Research\altaffilmark{1}, PO Box 3000, Boulder, CO 80307 USA}
\altaffiltext{1}{NCAR is sponsored by the National Science Foundation.}

\author{Ada Ortiz\altaffilmark{2}}
\affil{High Altitude Observatory, National Center for Atmospheric
Research\altaffilmark{1}, PO Box 3000, Boulder, CO 80307 USA}
\altaffiltext{2}{Current address: Departament d'Astronomia i Meteorologia, Universitat de Barcelona, Barcelona,
Spain, E-08028}

\author{Randle W. Meisner}
\affil{Laboratory for Atmospheric and Space Physics,
University of Colorado, Boulder, CO 80303 USA}

\begin{abstract}

We have examined images from the Precision Solar Photometric Telescope (PSPT) at the Mauna
Loa Solar Observatory (MLSO) in search of latitudinal variation in the solar photospheric
intensity. Along with the expected brightening of the solar activity belts, we have found
a weak  enhancement of the mean continuum intensity at polar latitudes (continuum
intensity enhancement $\sim0.1 - 0.2\%$ corresponding to a brightness temperature
enhancement of $\sim2.5{\rm K}$). This appears to be thermal in origin and not due to a
polar accumulation of weak magnetic elements, with both the continuum and CaIIK intensity
distributions shifted towards higher values with little change in shape from their
mid-latitude distributions. Since the enhancement is of low spatial frequency and of very
small amplitude  it is difficult to separate from systematic instrumental and processing
errors. We provide a  thorough discussion of these and conclude that the measurement
captures real solar latitudinal intensity variations.

\end{abstract}

\keywords{Sun: activity --- Sun: faculae, plages --- Sun: magnetic fields --- Sun:
photosphere --- Sun: rotation}

\section{Introduction}

Latitudinal variations in the thermal structure of the solar convective envelope have been
implied or required by theories ranging from non-general-relativistic gravity \citep[]{dic64}
to those explaining the solar differential rotation \citep[e.g.][]{kit95,dur99,rem05,mie06},
meridional circulation, and torsional oscillations \citep[]{spr03}.
Models of the solar differential rotation and meridional circulation reproduce the observed
helioseismic rotation profiles \citep[]{thom96,sch98} best
when baroclinicity is introduced via a latitudinal 
temperature gradient in region of 
the solar tachocline which spreads into the convection zone proper
by turbulent convection \citep[]{rem05,mie06}.  
While it is estimated from such models that in the solar photosphere
the pole may be as much as 
a few degrees warmer and the mid-latitudes a couple of degrees cooler than the equatorial region \citep[e.g][]{bru02,mie06},
these estimates are model dependent and
measurements have proven difficult. 
The difficulties arise because
of the intrinsically low amplitude thermal signal expected
and the presence of magnetic structures which introduce small scale intensity fluctuations of 
comparable or greater amplitude.  Thus, previous observations have yielded 
wide ranging results (Table~\ref{table1}).

Here we examine full disk images from the
Precision Solar Photometric Telescope (PSPT)  at the Mauna Loa Solar Observatory (MLSO).  
The telescope should by design readily allow measurement of latitudinal temperature variations 
if they are a few degrees in
magnitude.
In \S2 below we review the telescope capabilities and data processing techniques necessary to 
realize those.  In \S3 we present the results obtained 
and in \S4 discuss the uncertainties in these 
due to random and systematic errors.  
Finally, we conclude in \S5 by putting our methods and
results in the context of previous measurements.

\section{The PSPT data}

The PSPT at the MLSO is a small (15 cm) refracting telescope designed for photometric
solar observations with the aim
of achieving an unprecedented 0.1\% pixel-to-pixel relative photometric precision. It
acquires full-disk solar images on a $2048\times2048$ CCD array ($\sim1\arcsec$/pixel) in three wavelength bands: blue
continuum ($409.412$ FWHM $0.267$nm), red continuum ($607.095$ FWHM $0.458$), and CaIIK ($393.415$ FWHM
$0.273$nm)\footnote{Two additional narrow band CaIIK filters have recently been added to the telescope, one at line-center and 
the other in the red wing, but these are not
used in this study.}.  The red continuum
band is quite clear of absorption lines while the blue continuum is not (profiles superimposed on reference
spectra are available on the PSPT web site). The quiet-Sun formation heights (relative to optical depth unity at
500 nm and over the FWHM of the filter band) for these wavelengths are approximately \citep[]{ras01}: blue continuum, -10 to 35 km;
red continuum, 0 to 50 km; and CaIIK line, 900 to 1660 km (line core) and 0 to 260 km (line wings). The PSPT
concept and prototype are described in more detail in \citet{ck94}.

\subsection{Gain correction}

\begin{figure}
\includegraphics[scale=0.6]{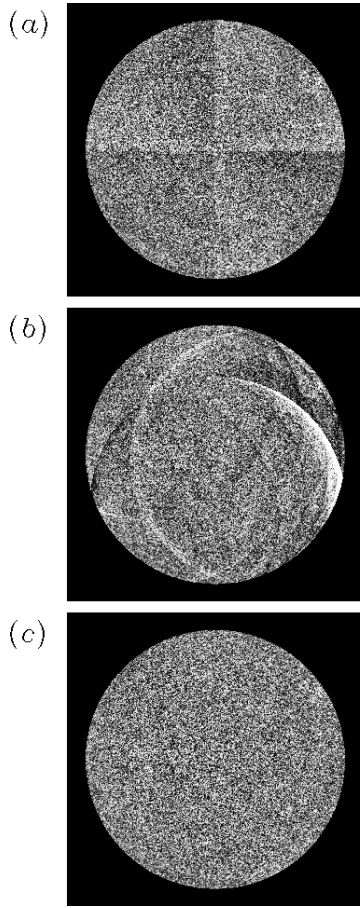}
\caption{Examples of failed ($a$ and $b$) and successful ($c$) 
gain correction attempts, illustrated with
histogram-equalized contrast images.  Note that 
the failed attempts chosen for this illustration have especially large and 
obvious problems which can be seen even in unequalized high contrast images.  
This is not usually the case, and 
most image defects remaining after `flat-fielding' are of much lower 
amplitude and can only be faintly seen after
histogram equalization of the image.  
No MLSO PSPT data is considered acceptable if the amplifier induced quadrant
structure is at all apparent in histogram equalized images. Every attempt 
is made to arrive at a contrast image showing no residual arc structure 
introduced by the~\citet{kll91} algorithm. \label{fig0}}
\end{figure}

The PSPT camera readout occurs in quadrants through four amplifiers of 
differing gain.
In order to achieve in data images the photometric precision promised 
by the instrument design, these and
other spatial variations in the detector must be
determined to an equivalent accuracy.  We accomplish this 
using a version of the algorithm of
\citet{kll91} applied to sixteen offset images of the Sun recorded once per day.
By assuming that the Sun is unchanged between each of these offset images and 
comparing both pixel intensities at identical image locations and pixel
gains at identical detector locations, 
the procedure captures a best fit spatial mapping to the differences between the offset 
image intensities which when corrected would minimize their differences.  
The procedure is sensitive to a variety of contributions,
including, those fixed on the detector plane
(such as pixel to pixel gain variation or the 
amplifier induced quadrants), those resulting from variation 
in the offset image light path (such as 
filter density variations), those resulting from image evolution 
during the acquisition sequence (seeing variations), and any others which are 
not stationary with respect to the solar image.  The sensitivity 
to variation in atmospheric seeing during the course of the
sequence is particularly troublesome, and a usable 
`flat-field' is not obtained daily. Figures~\ref{fig0}$a$ and $b$ display,
with histogram equalization of the intensity~\citep[e.g.][]{rus99}, 
contrast images (see below) which are produced from the raw data using
particularly poor flat-fields.  The gain correction in ($a$) failed
to remove the 
amplifier induced quadrants and in ($b$)
introduced arc shaped defects related to the
offset image positions in the flat-field algorithm. 
Such flat-fields would never be used in
the image processing (Figure~\ref{fig0}$c$ displays the image 
as actually processed for this study).
All PSPT images from the MLSO are inspected
individually using histogram equalization of the contrast
intensity and, if necessary, reprocessed with a flat-field 
from a neighboring date until
a gain correction is made which leaves no visible quadrant 
structure and no or few algorithm related defects in the image. 
This kind of inspection is capable of identifying defects remaining in 
individual images to very low amplitudes 
(below 0.1\%),
but, since the flat-field image offset
positions are a fixed sequence on the detector, defects remaining in the images 
below human visual acuity 
are the largest known source of systematic error in the data. We discuss these in detail in 
\S4.

\subsection{Center-to-limb variation}

While the solar center-to-limb intensity variation 
and its possible temporal variability 
has intrinsic importance, reflecting the mean thermodynamic structure of the outer layers 
of the star, 
here we are interested in spatial variations (in particular latitudinal variation) about this
underlying azimuthally invariant limb-darkening profile,
and so must also measure it with precision.
We employ a procedure 
which identifies quiet sun pixels in the image and performs a 
simultaneous least-squares fit to the intensity of these
to determine the coefficients of a truncated series of Legendre
polynomials in radius $P_m(r)$
and Fourier modes in central angle:
$$
I_0(r,\theta)=\sum_n\sum_m\left[A_{nm}P_m(r)\cos(n\theta)+B_{nm}P_m(r)\sin(n\theta)\right]
$$
truncated to $m\le 6$ and $n\le 2$,
where $r$ and $\theta$ are polar coordinates with the
origin at the solar disk center.
Quiet sun pixels are taken as those 
within $\pm2.5\sigma$, in red and blue continuum, or $-2.5<\sigma<0.0$, in
CaIIK images, of the median intensity in about seventy-five
equal area annuli centered around the solar disk center (the annulus area is fixed so the
number employed is slightly dependent on filter band and time of year).  
The orthogonality of the fitted 
functions allows capture, with little mixing, of both 
the solar center-to-limb variation and any residual dipolar (linear gradient) 
or quadrupolar (quadrant structure) artifacts present in the image.  
We note that, because the Fourier phase at a given radius is
independent from that at another, this formalism allows 
for the removal of a warped gradient or quadrant
surface (the Fourier fit can rotate with
azimuth at a given $r$). This is a powerful 
tool in removing large scale image defects (see \S4).
We also note that the use of the contrast images $I_c=(I-I_0)/I_0$,
where $I$ is the intensity measured at each pixel and 
$I_{0}$ is the fit to the solar center-to-limb
variation and residual defects, in this study implies that the intensity measurements reported here are  relative
to a  particular center-to-limb profile removed.  That profile was calculated for a particular subset of quiet Sun
pixels in each image (defined above), and other pixel subsets in the resulting contrast image will show
center-to-limb brightening or darkening depending on their activity level in relation to those used to compute
the center-to-limb function removed.  

\subsection{Data selection and alignment}

We have selected fifty six of the best contrast image triplets from the MLSO 
PSPT image archive, spanning the period
from March 2005 to July 2006. The complete list of selected image triplets is given in Table~\ref{table2}. 
Image quality was determined using two criterion: quality of the gain correction (\S2.1) and atmospheric seeing
conditions. Seeing was estimated based on a measure of the red continuum solar limb width as determined from
a Gaussian profile fit
to intensity values limb-ward of the inflection point in the four grid aligned directions from the solar disk
center.  Only images with an average red continuum Gaussian FWHM of less than 2.5 pixels (limb width of 1.25 pixels) were used in
this study.

Following selection, each image was resized, rotated, and aligned to a common reference
grid in which the solar P0 and B0 angles are zero, i.e., the vertical image axes were
aligned with the solar North-South. 
While these operations
require interpolation (bilinear used) and thus introduce error, we have minimized
the total impact by combining the procedures into a single interpolative step in heliographic 
coordinates.

\subsection{Activity masks}

\begin{figure*}
\begin{center}
\includegraphics[scale=0.9]{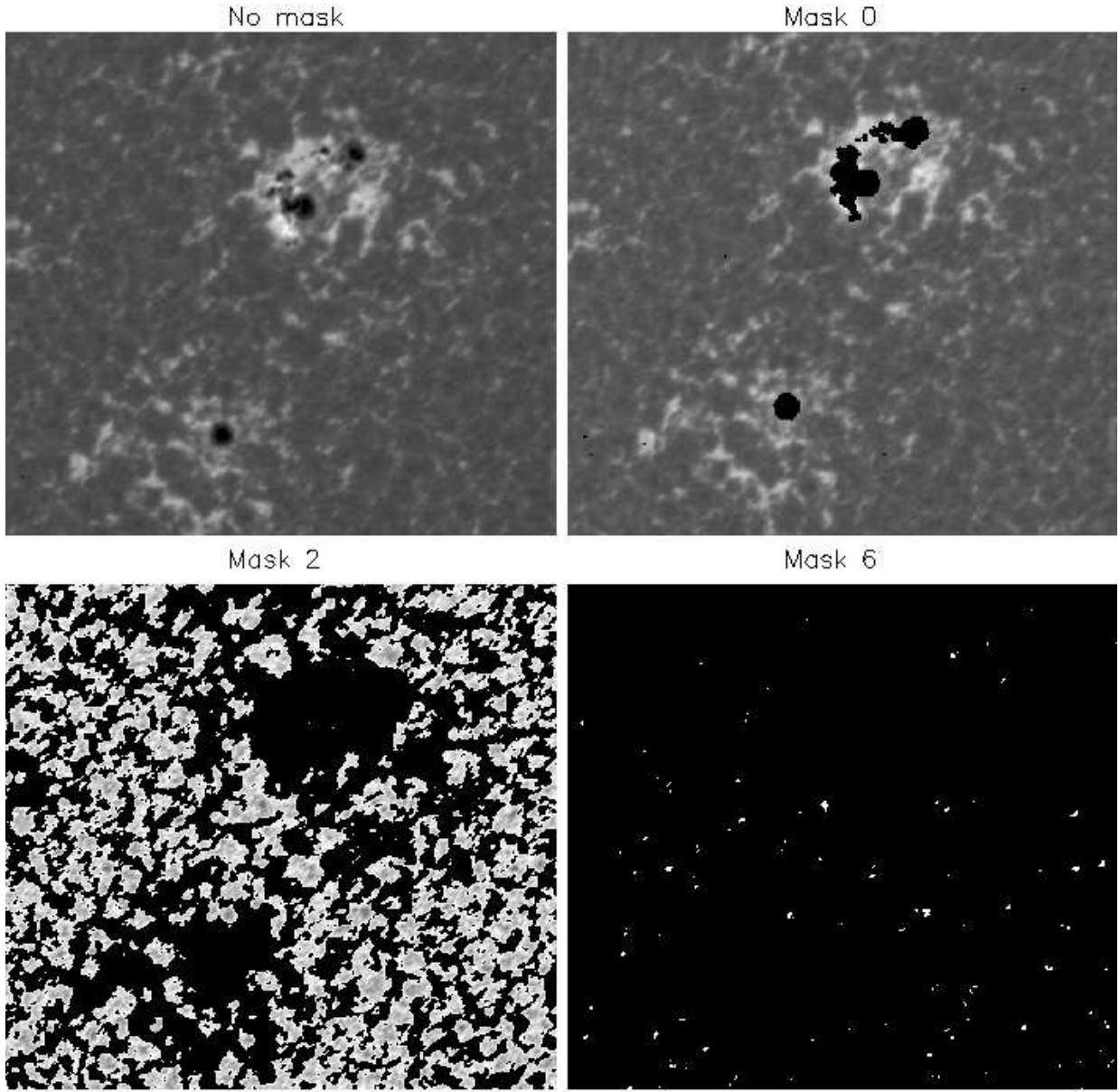}
\caption{Example applications of CaIIK masks to the same 
sub-image. Intensity thresholds are used to mask out activity.
While Mask 2 eliminates most plage and network elements retaining 
about 50\% of the pixels, Mask 6 retains only 0.15\% of the pixels
masking out all
but the darkest cell
centers.\label{fig1}}
\end{center}
\end{figure*}

Once aligned, the images can be used to construct averages and masks. 
Pixel selection by masking (here we define masking as
excluding from analysis)
is based on using CaIIK intensity as a surrogate for the magnetic flux density
\citep[e.g.][]{sku75,schetal89,har99,ras03,ort05}. After sunspots and pores are masked out
by thresholding the red continuum contrast intensity (keeping $I_{R}>-0.05$), small magnetic
elements show a strong positive
correlation between CaIIK intensity and magnetic flux density, even
down to very low intensity values \citep[]{ras03}. Here we apply increasingly severe masks to 
the images, retaining only pixels in the analysis whose CaIIK intensity falls below the
mask value. Table~\ref{table3} indicates the mask intensity values used in this study, the
approximate corresponding 
magnetic flux density \citep[from][]{ras03,ort05}, and the average (over the image set)
percentage of pixels retained
by each.
Masks range from very unrestrictive (M0 retains all magnetic activity except
sunspots and pores) to very restrictive (application of mask M6 or M7 yields images
which retain only the very darkest CaIIK network cell centers). 
Figure~\ref{fig1}
illustrates the effect of applying masks M0, M2, and M6 to a subregion of a typical CaIIK
image.

\begin{figure*}
\begin{center}
\includegraphics[scale=0.9]{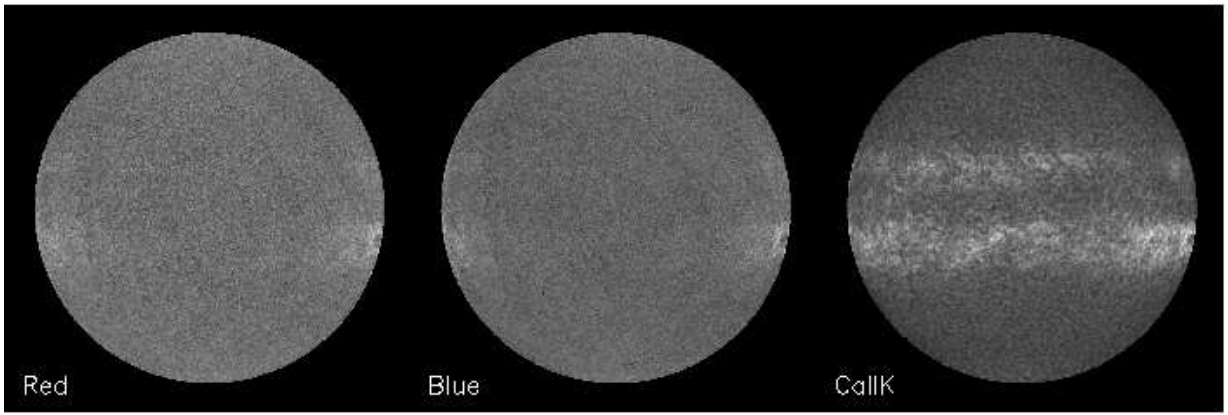}
\caption{Average contrast image after application of mask M0 to the images
at each of the three wavelengths.  Magnetic activity
in the active belts shows hemispherical asymmetry during the declining phase of Cycle 23.  Also faintly visible is the polar brightening which is the subject of this paper.  Note that the polar brightening measured is not that due to the 
extreme limb points, which are somewhat unreliable due to reduced image contributions as a result of the B0 correction.  
The bulk of the analysis presented in this paper
uses pixels from annuli well inside the limb (see Figure~\ref{syse}$a$).
\label{fig2}}
\end{center}
\end{figure*}

We applied the eight masks based on CaIIK contrast values
to the 56 contrast images at all wavelengths 
and averaged the unmasked pixels to produce
mean contrast images at each wavelength and activity level. 
Examples are shown in Figure~\ref{fig2}, 
where the average image (after mask M0 application) at
each wavelength is displayed. The average activity over the nearly 17 months of observation
is apparent as bright latitudinal bands across the CaIIK image disk and as limb-brightened 
faculae in the
continuum observations. 
As has been previously reported for the declining phase of Cycle 23 \citep[]{kna04,zat06},
hemispherical asymmetry in magnetic activity is clearly evident in these average images
(it is also seen in a more uniform sampling of the observation period).
The continuum images also show a hint of the polar brightening which is the subject of this
paper. 

In next section we present results obtained largely from the application of masks M0 and M2
which separate plage/network from non-network contributions. Mask M0 removes only sunspots
and pores from the images while M2 masks out most network elements (Figure~\ref{fig1}). 
Preliminary results using the more severe mask M6 to remove all but the contributions from
the darkest cell centers (these perhaps representative of purely thermal 
emission from the non-magnetized photosphere) are also presented.

\section{Results}

Using the techniques described we have examined the latitudinal variation of the continuum
photospheric intensity in PSPT images. For each activity mask and wavelength, we have
longitudinally averaged the 
mean contrast image over $2\degr$ wide latitudinal bins. 
The results are shown in Figure~\ref{fig3}.  To be
meaningfully interpreted, the variation with latitude must be compared with the
longitudinally averaged center-to-limb variation (CLV) of the 
mean contrast image (as discussed in \S2.1 the unmasked pixels show a CLV relative to that already removed in making the contrast image). This
is plotted in Figure~\ref{fig3} with solid curves. If
no systematic latitudinal variation in mean intensity 
existed on the Sun then the measured longitudinally
averaged mean contrast would be the same 
as the longitudinally averaged CLV. This is 
not the case.  Significant 
(see \S4 for discussion of noise properties) variation
in the longitudinally averaged 
intensity is observed with latitude.  

Images weakly masked by M0 to remove only sunspot and pore contributions
(top row of Figure~\ref{fig3}) clearly show enhanced intensity at active region 
latitudes, as was already apparent in the full disk images of Figure~\ref{fig2}. 
The peak contrast intensities occur 
near $\pm12\degr$ latitiude at all wavelengths, with greater enhancement in the 
southern hemisphere than the northern. 
Additionally, the images masked by either the M0 or the
M2 mask (middle row of Figure~\ref{fig3}) show evidence of polar brightening. 
Mask M2 eliminates most magnetic activity over
the solar disk (only pixels whose CaIIK intensities correspond to magnetic flux densities of 
less than a couple gauss
have been kept), and after application of this mask
the images show an average $\sim0.1\%$ enhancement in contrast
in the polar region in both
the red and blue continuum bands, slightly larger in the blue.  
The CaIIK contrast for this mask value shows no polar enhancement 
in the southern hemisphere but some in the northern. This is not yet understood.

The observed polar continuum enhancement is likely not due to the presence of
polar faculae. While these may contribute to the unmasked profiles of M0, 
they are eliminated by  M2 mask as their
CaIIK intensity likely exceeds the $I_{K,max}$ threshold.
This assertion is supported by the 
pixel intensity distributions (Figure~\ref{figh}) which show that the polar and
mid-latitude distributions have nearly identical shapes, lacking the bright tail of the 
equatorial distributions.   The polar distributions at all wavelengths are
shifted as a whole to higher intensity values than 
their mid-latitude counterparts. This is
in agreement with flux measurements 
using the solar distortion telescope \citep[]{kld88}.  
The lack of any bright  
CaIIK distribution tail, combined with the 
shift toward higher intensity in
the continuum bands, 
suggests a thermal non-magnetic origin for the 
polar brightening.  That the
CaIIK distribution is also
shifted likely results from the fact that PSPT CaIIK
filter is sufficiently broad
to include a significant 
photospheric contribution.

Images masked by the even more severe M6 mask exclude all contributions but those from the
quietest CaIIK cell centers. Application of this mask is ideal for determination of the
latitudinal variation in the thermal structure of the solar photosphere, as virtually all
magnetic contributions should be negligible. Unfortunately, the small number of remaining
pixels makes this measure subject to large statistical uncertainty. 
The noise is largely
of solar origin (unresolved granulation) and could be addressed with more data. This may be 
worthwhile, as the preliminary results presented here (bottom row Figure~\ref{fig3}) suggest a
hint of the expected \citep[]{bru02,rem05,mie06}
increase in photospheric continuum intensity
in both the polar {\it and} equatorial regions as compared to the solar mid-latitudes.  If these increases are found to be
significant after analysis of additional data,
they would confirm a non-facular origin of the
equatorial brightening \citep[]{kld87, kld88},
as all facular contributions have likely been eliminated by
the severity of the M6 masking.

An alternative way of plotting the latitudinal intensity variations 
(to further distinguish them from those of the
CLV, and later in \S4 from systematic error) 
is to plot the contrast as a function of polar angle $\theta$ on the disk at fixed cosine
of the heliocentric angle $\mu$.  Since the sampling now occurs at fixed
$\mu$, the CLV contributes a constant offset to the measurement
(independent of $\theta$).
Figures~\ref{fig4}$a$ and $b$ present the continuum red and blue
contrast intensities in this way after applying masks M0 and M2, respectively. 
The average contrast values in a ring
between $0.3<\mu<0.45$ and for $2\degr$ bins in $\theta$ are plotted, 
scanning the solar limb counter-clockwise
from the west (right hand side of the image). 
In these plots the equator lies at $\theta=0$, $\pi$ and $2\pi$, 
while the poles lie
at $\pi/2$ and $3\pi/2$. 
The hemispherically asymmetric 
north and south activity belts are again apparent in Figure~\ref{fig4}$a$ as increases
in the red and blue continuum intensities either side of the equator. 
The polar enhancement is also prominent as
are the darker mid-latitudes. Figure~\ref{fig4}$b$ further confirms the polar 
enhancement.
After application of mask M2 the polar intensity is 
about 0.15\% larger than that at the equator in the red continuum  
($\sim0.2$\% greater 
at the pole than equator in the blue continuum).

\section{Random and Systematic Errors}

Random fluctuations in the PSPT images are largely granular in origin
and thus of high spatial frequency, follow a $\sqrt{n}$ reduction in
amplitude upon image averaging (when the images are 
separated by more than one granule lifetime), 
and are well represented by the error
bars shown in Figures~4 and 6.  Those error bars indicate 
the maximum rms fluctuations about the 
latitudinal or $\theta$ bin averages for any of the 
data points plotted. The variations with latitude 
or $\theta$ measured are thus significant with respect to small 
scale random noise.
Moreover, smaller subsets of the data images show the same underlying signal, 
albeit with larger amplitude fluctuations.

The known systematic errors in the data behave differently.  They are
of large spatial scale and persistent with
averaging, and
thus pose the greatest challenge to the measurement.
The dominant source of known systematic error is the 
presence of very low amplitude residual image defects
which reflect the offset positions used in the ~\citet{kll91} flat-field
algorithm (\S2.1).  Since their location is nearly fixed 
(somewhat smeared 
by the P0 and B0 correction during image alignment), they add
upon averaging and are faintly visible in the
continuum contrast images in Figure~\ref{fig3}.
This section
addresses and quantifies the level of uncertainty introduced by this
source of error.

As discussed in \S~2.1 the PSPT images are subject to
two consecutive image processing stages, the first aimed at removing detector
gain variations and the second, combined with the center-to-limb calculation,
intended to remove residual dipolar and quadrupolar defects. This
second step is necessary, not only to capture the faint defects described 
above, but also because the flat-field algorithm itself
is subject to an
intrinsic gradient ambiguity~\citep[]{kll91} and can thus introduce or fail to
remove linear gradients in the image. The gradient is measured via
the $n=1$ Fourier component of the CLV routine. We will see below that the 
$n=2$ mode proves useful in the further removal of the systematic defects.

In order to assess the efficacy of these techniques,
it is useful to apply them 
to those portions of the image which lie off of
the solar limb in the same way as they are applied to the solar disk itself.
This allows comparison between the systematic noise contributions to the
image and the measured signal,
since the signal should be absent off of the solar disk.
While the standard PSPT processing procedures correct 
for the detector gain variation across the entire field of view, they apply
the Legendre/Fourier center-to-limb/defect correction only 
on the solar disk. In this section
we thus employ a conceptually (but not mathematically) equivalent procedure
which can be readily extended off of the solar limb.

After masking sunspot pixels, we divide
the image by the median value of the intensity in
very narrow annuli (1000 equal width annuli employed)
centered on the solar disk center.  This 
very effectively removes the azimuthally invariant
center-to-limb variation on the disk, while 
performing an equivalent
correction off of it.  Fourier amplitudes are then calculated, fitting
the pixels within each annulus independently with the $n=0-2$ expansion.
The dipolar and quadrupolar 
defect correction is thus extended off of the solar disk.
Figure~\ref{syse}$a$ displays a single red continuum
image processed in this way after histogram equalization of the contrast
intensity.  While the large scale defects contributing to systematic error in 
the analysis are barely visible on the solar disk, where real physical intensity
fluctuations dominate the image and the error contrast is low, 
they are quite apparent 
as both residual quadrants and flat-field induced arc defects off the limb,
where their relative contrast is high due to the low mean intensity values.
Figures~\ref{syse}$b$ and $c$ display the mean contrast, 
after application of masks
M0 and M2 respectively, in two annuli (on the
disk {\it solid} curve, off the disk {\it dashed} curve) as a function of 
$\theta$.  
These curves are the averages of
those obtained from the individual images after shifting appropriately
for the solar P0 angle, and the off disk contrast has been scaled down
by the ratio of the median intensity in the two annuli.
The {\it solid} curves are equivalent to the red continuum plots shown in 
Figures~\ref{fig4}$a$ and $b$ but without the B0 correction, and 
we note the very close similarity to those despite this 
difference and that of the underlying CLV correction method.
What Figure~\ref{syse} displays in addition to Figures~\ref{fig4} is a measure
of the systematic processing errors off of the solar disk ({\it dashed} curves).
These are much smaller in magnitude than
the measured signal on the disk suggesting that the measurement is not 
dominated by noise.  

Since the measured polar brightening takes a $\cos(2\theta)$ form,
it is important to ask if the Fourier mode correction for image defects in 
fact spuriously introduces the `signal' being measured.  
Figures~\ref{syse2}$a$ and $b$ plot the average contrast 
(after mask M2 application)
as a function of theta after employing only the $n=0-1$ and $n=0-2$ corrections respectively.  The amplitude of the polar brightening is reduced, not enhanced
by the $2\theta$ correction.  It is also apparent that the $2\theta$ correction 
further reduces the systematic noise off of the solar disk 
({\it dashed} curves).  Moreover,  Figures~\ref{syse2}$a$ and $b$ plot 
separately the contributions, both on and off of the disk, made by images with 
P0$\ >10.0$ and P0$\ <-10.0$ ({\it dark} and {\it light} grey curves respectively). 
This is a robust measure of the error contributions to the signal since any 
real solar measurement should be stationary with respect to this distinction.
Without the $2\theta$ correction (Figure~\ref{syse2}$a$) the positive and 
negative P0 curves, from the data
both on and off of the disk, are consistently 
shifted with respect to one another.  After application of the 
$2\theta$ correction (Figure~\ref{syse2}$b$) the signals off of the solar disk 
remains consistently shifted, but those from on the solar disk do not.  
After the correction, 
the southern hemisphere signals for positive and
negative P0 values show no or
even slightly the wrong sense of shift with respect 
to one another, indicating that 
they are a property of the solar image and not a defect fixed
in the detector
plane.  The signal in the northern 
hemisphere still appears to have some noise contribution, 
but at a much reduced amplitude. This is consistent with our experience with 
the images, which tend to show more defects in the northern hemisphere in 
response to the fixed flat-field sequence.
We note that this test was
also performed on subsets of the data,
after selection of just the very best images, and
while the random noise levels were higher, the polar brightening always stood
out from the systematic error. Moreover, annuli of smaller radius showed 
the same signal, albeit of decreasing amplitude as expected.
We conclude that the Fourier/Legendre CLV 
processing allows measurement of solar polar brightening at the very limits of
the PSPT capabilities.

\section{Conclusion}

We have measured an enhancement of the photospheric continuum intensity 
in the solar polar regions using PSPT
images. We have carefully examined the properties of the signal and 
believe it to be of solar origin.  Moreover, the properties of the intensity
distributions suggest that the signal is thermal not magnetic in origin, 
although contributions 
from very weak network elements associated 
with the pole-ward extending branch of the solar cycle 
\citep[e.g.][]{how81,mul84,she06} cannot be ruled out 
without processing significantly more data and employing 
masking of greater severity.
This is planned, but limitations of the current ground based solar photometric
observations may make a decisive measurement difficult. Space based 
imaging photometry would be ideal for this and other intriguing 
problems.  This is particularly true if solar cycle 
dependencies are to be uncovered.

\begin{figure*}
\begin{center}
\includegraphics[angle=90,scale=0.7]{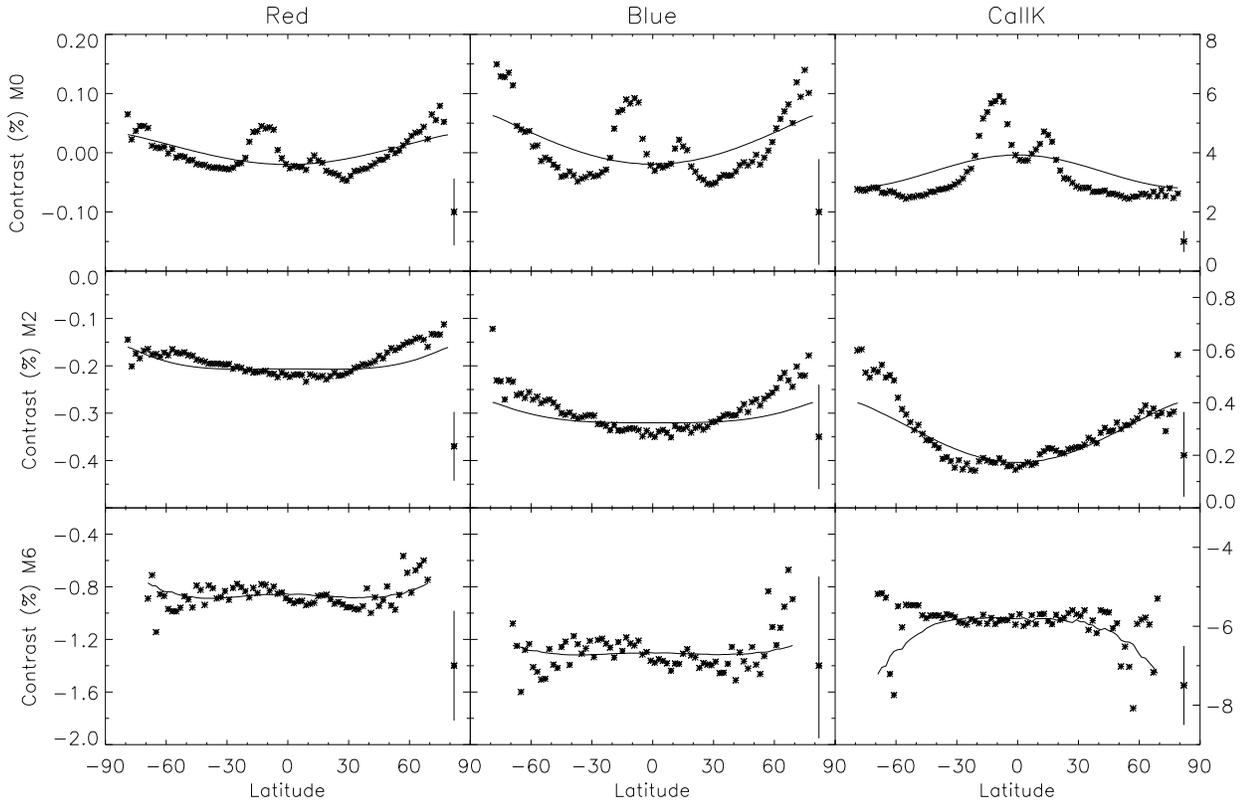}
\vspace{0.5cm}
\caption{Intensity contrast as a function of latitude for the red, blue and CaIIK
wavelengths after application of 
activity masks, M0, M2 and M6. The vertical scale on the {\it left}
is common for both the red and blue plots.  That for CaIIK is indicated 
on the {\it right}.
Solid curves plot the same longitudinal average of the 
center-to-limb variation as a function of latitude.
Vertical bars indicate the largest error estimated from the
rms fluctuations in each latitudinal bin.
\label{fig3}}
\end{center}
\end{figure*}

\begin{figure*}
\centering
\includegraphics[scale=0.75]{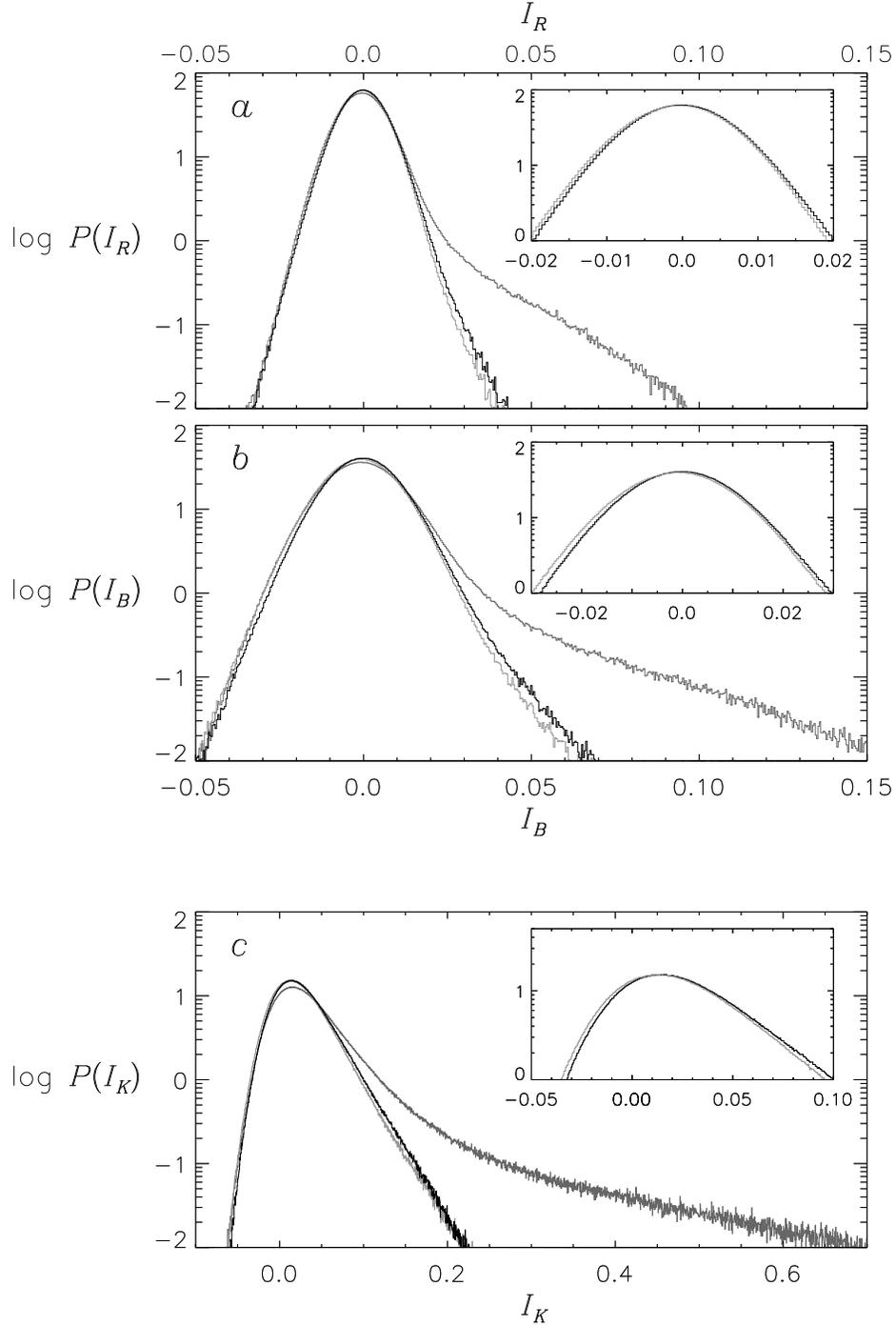}
\caption{Contrast distributions computed for each of the 
unmasked images, normalized so that integrated probability is unity 
and averaged over the 56 individual images in each wavelength. 
The distributions are those of pixels from an annulus 
between $\mu=0.3$ and $\mu=0.45$ and separated into polar contributions  
({\it black} curves), equatorial contributions ({\it dark grey} curves),
and mid-latitude contributions ({\it light grey} curves)
as defined in Figure~\ref{fig4}.  Insets show close-ups of the 
mid-latitude and polar distribution 
peaks.
\label{figh}}
\end{figure*}

\begin{figure*}
\centering
\includegraphics[width=8.cm,height=8.cm]{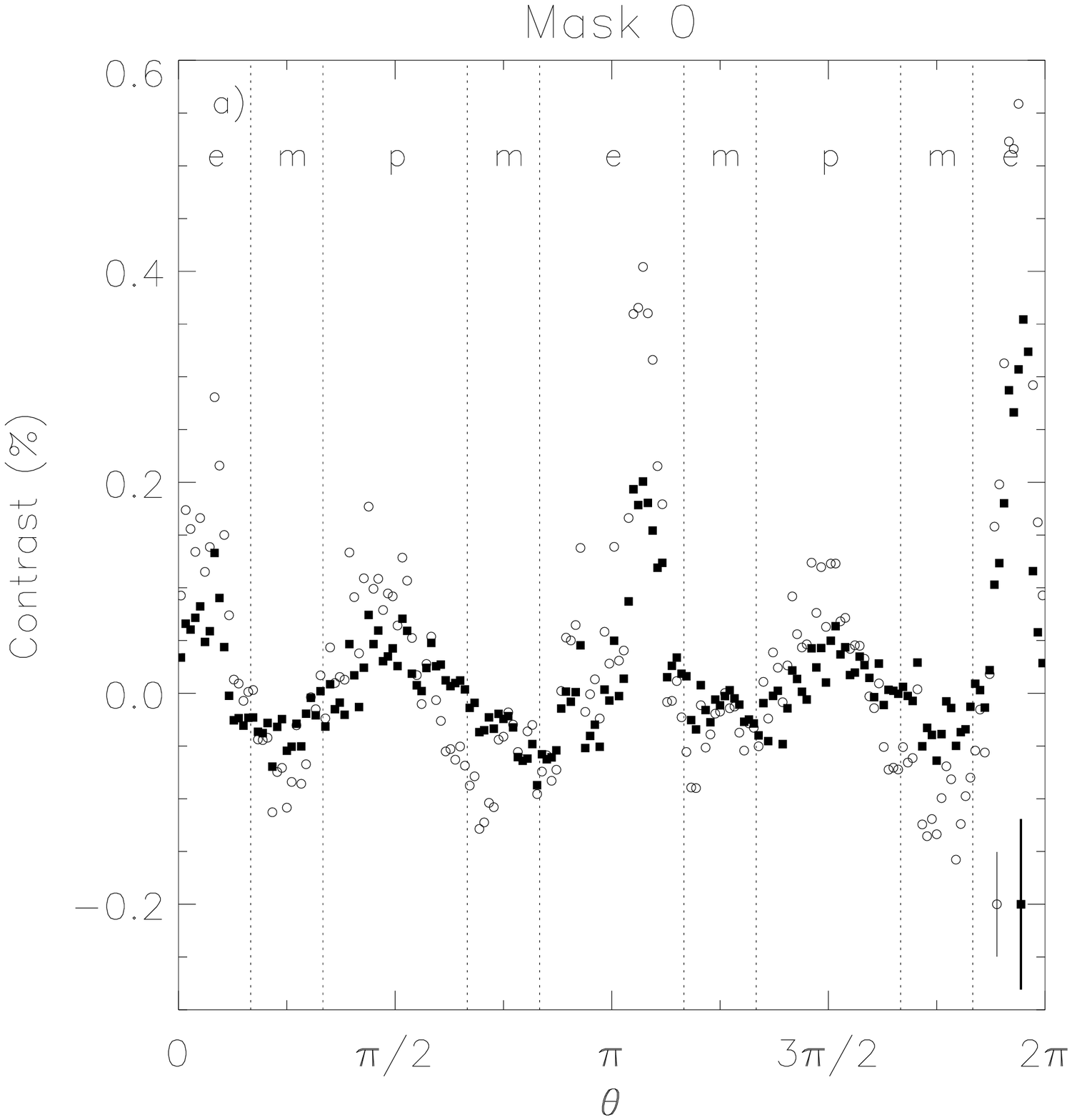}
\includegraphics[width=8.cm,height=8.cm]{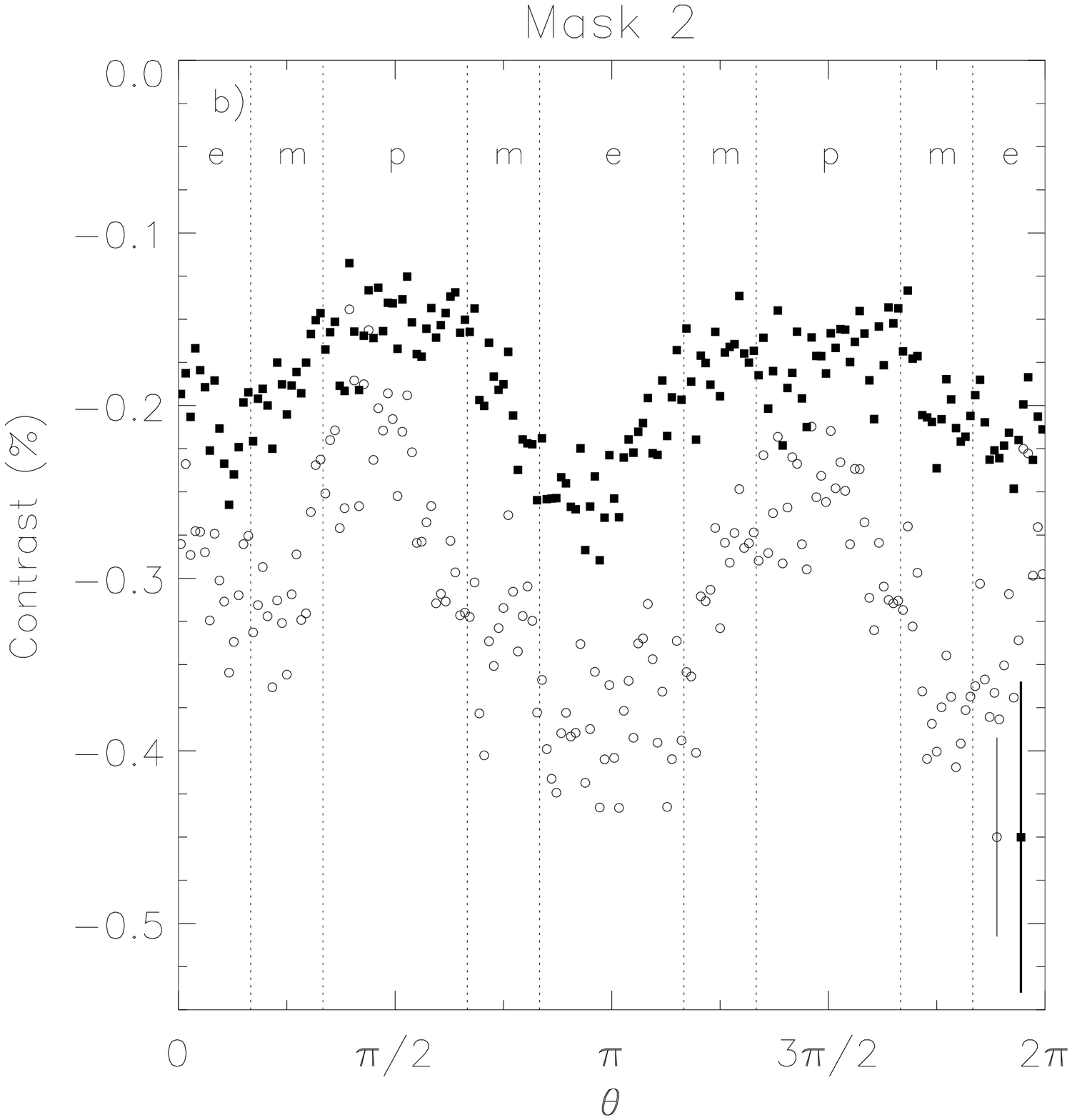}
\caption{Red (filled squares) and blue (open circles) contrasts 
in the average M0 and M2 masked images as functions of
polar angle $\theta$ around the annulus
$0.3<\mu<0.45$.
Three sectors in $\theta$, equator,
mid-latitudes, and poles, are marked by the letters {\it e}, {\it m}, 
and {\it p}.
Vertical bars indicate the amplitude of the maximum 
rms fluctuations in any bin.
\label{fig4}}
\end{figure*}

\begin{figure*}
\centering
\includegraphics[scale=0.9]{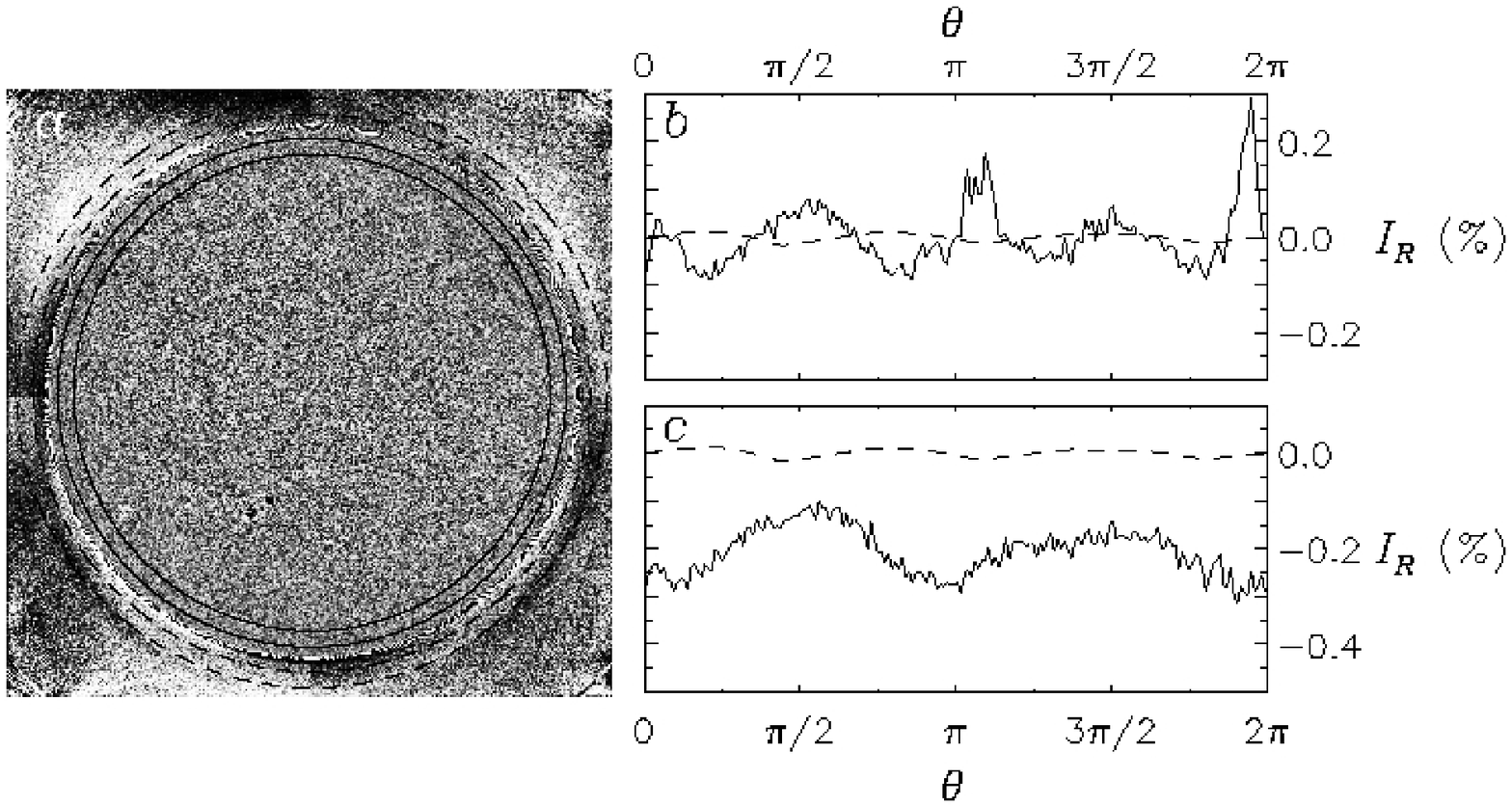}
\caption{In ($a$), a histogram equalized red contrast image from 
14 August 2005 with the CLV and residual defect
processing applied to the full field of view (see text).
In ($b$) and $c$), average (over all images and in $2\degr$ bins in $\theta$ 
for the $0.3<\mu<0.45$ annulus) 
contrast as a function of $\theta$
in two annuli (shown in ($a$)), one off of ({\it dashed} curve) and one on
({\it solid} curve) the solar disk, after application of masks M0 (in $b$) and M2 (in $c$).
\label{syse}}
\end{figure*}

\begin{figure*}
\centering
\includegraphics[scale=1.0]{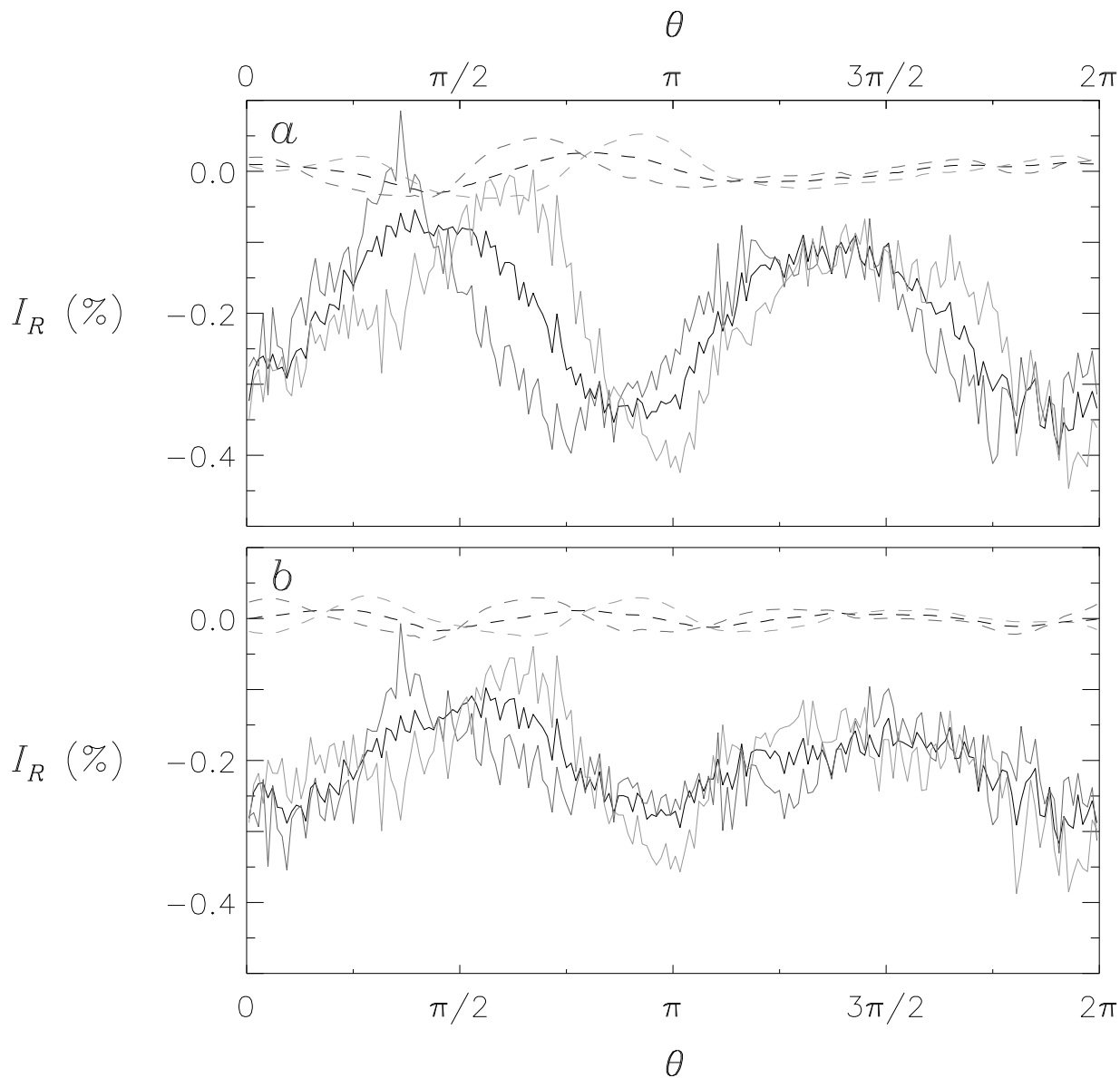}
\caption{Average (over all images and in $2\degr$ bins in $\theta$
for the $0.3<\mu<0.45$ annulus)
contrast as a function of $\theta$
in two annuli, one off of ({\it dashed black} curve) and one on
({\it solid black} curve) the solar disk (see Figures~\ref{syse}$a$), 
after application of mask M2.  In ($a$) only the $n=0-1$ Fourier modes have 
been included in the defect correction, in ($b$) $n=0,2$ modes are included.
{\it Dark} and {\it light} grey curves plot the same quantities for 
images with P0$\ >10$ and 
P0$\ <-10$ respectively.
\label{syse2}}
\end{figure*}

Such measurements would help constrain global solar models by adding a 
thermodynamic constraint to the dynamical constraints 
currently provided by 
helioseismology \citep[]{kld88}.  
Current numerical simulations seem to show little 
indication of a local temperature minimum at the equator, as 
our surface measurement implies,
but rather display a mid-latitude minimum with a
weak local maximum at the equator and stronger one at the pole.  This is true
even when a monotonically decreasing pole to equator profile 
is imposed in the solar
tachocline \citep[]{rem05,mie06}.
While we find some hint of equatorial brightening
after application of
our most restrictive activity masks (bottom row Figure~\ref{fig3}), 
it has extremely low amplitude and the measurement is extremely
uncertain.  

Our efforts can be distinquished from previous work in several
ways. Early work (references $1-7$ in Table~1) struggled with
instrumental precision, systematic error,
and detector gain correction. We too 
have taken pains to address these. 
More importantly, no distinction was made in that early work
between magnetic and thermal contributions to the signal, 
other than perhaps qualitative selection of nonactive regions.
We have emphasized this distinction by applying a series of
masks based on CaIIK intensity to the images before analysis.  
Previous efforts by Kuhn 
et al. (references $8-10$ in Table~$10$) have, for the most part,
also attempted to separate quiet Sun and facular contributions
to the latitudinal signal. The approach in those studies differed
from ours in that the facular contribution to the 
latitudinal variation was estimated statistically
based on models of the quiet Sun and facular 
color and intensity 
distributions.
Interestingly, the 
pole/mid-latitiude temperature contrast measured in
those studies is in 
good agreement with our measurement. 
The previous authors, however, reported an
additional low latitude brightness enhancement, only seeen 
very marginally, if at all, in our study at the 
most severe masking levels.
We note, that the one measurement of solar
latitudinal intensity variation available using spacecraft data
\citep[]{kuh98} did not remove the facular contribution during
analysis. 
The latitudinal temperature profile obtained in that work
looks very similar to our Figure~$6a$, also unmasked. 

Finally, we caution that our analysis does not 
completely eliminate possible contributions to the solar
photospheric intensity 
from an abundance of unresolved
magnetic elements of very low magnetic flux density.
Correlations between magnetic flux density and CaIIK intensity 
extend to very low values, making masking
based on these images possible, but also suggesting continued 
contamination of the continuum measurements 
at very low contrasts (Rast 2003). 
Unambiguous measurement of 
a strictly thermal signal in the solar photosphere may
require space-based full-disk photometric imaging 
at high precision and resolution over many wavelengths,
combined with
detailed modeling of the radiation field. Even that effort 
may be stymied if in fact there is no quiet Sun.
 
\acknowledgments

Special thanks to S. Criscuoli, J. Harder, T. Holzer, and M. Rempel.
AO acknowledges partial support from grant AYA2004-03022 of the PNAA programme 
of the Spanish Ministry of Science and Technology.

\clearpage
\begin{deluxetable}{rrrr}
\tablecolumns{4} \tabletypesize{\scriptsize} \tablewidth{0pc} \tablecaption{Some past observations of latitudinal variation in the solar
photospheric temperature\label{table1}} \tablehead{
\colhead{Reference} & \colhead{$T_{pole}-T_{equator}$} & \colhead{Method} & \colhead{Active regions or quiet Sun} \\
\colhead{ } & \colhead{(K)}   & \colhead{ }    & \colhead{} } \startdata
1 & $< 3$ & limb flux measurements & no distinction \\
2 & 0 \tiny{(within errors)} & equivalent width of Fraunhofer lines at limb & no distinction \\
3 & 1.5 $\pm$ 0.6 & spectrograph scans of CLV & plage region excluded \\
4 & 3 $\pm$ 3 & Mg I spectra  at 50 disk positions & quiet Sun \\
5 & 0 $\pm$ 0.6\% & equivalent width of Fraunhofer lines at limb & avoid active regions \\
6 & $< 1$ & IR opacity minimum, grating monochrometer CLV scans & no distinction \\
7 & 19 \tiny{(South pole)} & continuum (501.15nm) CLV scans & quiet Sun \\
7 & -19 \tiny{(North pole)} & continuum (501.15nm) CLV scans & quiet Sun \\
8 & 1.2 $\pm$ 0.2 & limb flux measurements & faculae rejection \\
9 & 1.5 & limb flux measurements & faculae rejection \\
10 & 1.5 & MDI images, limb shape and brightness & no distinction \\
\enddata
\tablecomments{References: (1) \citet{dic67}; (2) \citet{cac70}; (3) \citet{ac72}; (4) \citet{rut73}; (5) \citet{fal74}; (6) \citet{kou77};
(7) \citet{wit78}; (8) \citet[][1987]{kld85}; (9) \citet{kld88}; (10) \citet{kuh98}}
\end{deluxetable}

\clearpage
\begin{deluxetable}{rrrr}
\tablecolumns{4}
\tabletypesize{\scriptsize}
\tablewidth{0pc}
\tablecaption{Census of the observed triplets \label{table2}}
\tablehead{
\colhead{Date} & \colhead{Hour} & \colhead{Date} & \colhead{Hour} \\
\colhead{ } & \colhead{(UT)}   & \colhead{ }    & \colhead{(UT)} 
}
\startdata
07 Mar 2005 & 18:50 & 19 Oct 2005 & 17:50 \\
17 Mar 2005 & 17:20 & 25 Oct 2005 & 18:30 \\
23 Mar 2005 & 17:50 & 25 Oct 2005 & 18:50 \\
12 Apr 2005 & 17:40 & 15 Nov 2005 & 18:30 \\
13 Apr 2005 & 17:12 & 18 Nov 2005 & 18:02 \\
14 Apr 2005 & 17:12 & 19 Nov 2005 & 18:02 \\
21 Apr 2005 & 18:20 & 30 Nov 2005 & 18:12 \\
25 Jun 2005 & 17:12 & 30 Dec 2005 & 18:12 \\
27 Jun 2005 & 16:50 & 31 Dec 2005 & 18:12 \\
27 Jun 2005 & 17:50 & 05 Jan 2006 & 18:20 \\
06 Jul 2005 & 17:02 & 18 Jan 2006 & 18:30 \\
20 Jul 2005 & 18:50 & 02 Feb 2006 & 19:02 \\
11 Aug 2005 & 17:02 & 06 Feb 2006 & 18:20 \\
14 Aug 2005 & 17:12 & 06 Feb 2006 & 19:02 \\
16 Aug 2005 & 17:02 & 08 Feb 2006 & 18:20 \\
17 Aug 2005 & 17:30 & 09 Feb 2006 & 18:50 \\
21 Aug 2005 & 17:30 & 28 Mar 2006 & 19:12 \\
22 Aug 2005 & 17:12 & 07 Apr 2006 & 18:12 \\
23 Aug 2005 & 17:30 & 14 Apr 2006 & 18:30 \\
23 Aug 2005 & 17:50 & 29 Apr 2006 & 17:12 \\
24 Aug 2005 & 17:12 & 03 May 2006 & 17:12 \\
07 Sep 2005 & 17:12 & 04 May 2006 & 16:50 \\
17 Sep 2005 & 17:20 & 04 May 2006 & 17:12 \\
23 Sep 2005 & 17:40 & 18 May 2006 & 17:20 \\
06 Oct 2005 & 18:12 & 18 Jun 2006 & 18:12 \\
06 Oct 2005 & 18:30 & 20 Jun 2006 & 17:12 \\
14 Oct 2005 & 18:02 & 20 Jun 2006 & 17:50 \\
18 Oct 2005 & 18:12 & 14 Jul 2006 & 17:02 \\
\enddata
\tablecomments{
Observation times indicated are approximate.  Each image is an average of 30 exposures
aimed at minimizing 
brightness variations due to solar oscillations.  The three filter observations are interleaved 
(three sequences of 5 blue, 5 CaIIK, 5 red, 5 red, 5 CaIIK, and 5 blue are taken over the course of 
about 5 minutes) so that the resulting images at each of the three wavelengths
may be considered 
`simultaneous'.  The imaging begins
about three minutes (during which time dark calibrations are made)
after the observation 
time quoted above and used in the image filenames.  Actual start and end
times of the image sequence can be found in the image headers.
}
\end{deluxetable}

\clearpage
\begin{deluxetable}{crrrr}
\tablecolumns{5}
\tablewidth{0pc}
\tablecaption{Definition of the CaIIK masks \label{table3}}
\tablehead{
\colhead{Mask} & \colhead{$I_{R,{\rm min}}$} & \colhead{$I_{K,{\rm max}}$} & \colhead{$B_{\rm max}$} & \colhead{Pixels retained} \\
\colhead{ } & \colhead{ } & \colhead{ } & \colhead{(G)} & \colhead{(\%)}}
\startdata
M0 & -0.05 & \nodata & \nodata & 99.90 \\
M1 & -0.05 & 0.150 & 29.0 & 97.00 \\
M2 & -0.05 & 0.025 & 2.5 & 50.00 \\
M3 & -0.05 & -0.010 & 0.8 & 10.00 \\
M4 & -0.05 & -0.020 & 0.7 & 5.00 \\
M5 & -0.05 & -0.030 & 0.6 & 2.00 \\
M6 & -0.05 & -0.050 & 0.45 & 0.15 \\
M7 & -0.05 & -0.060 & 0.4 & 0.04 \\
\enddata
\tablecomments{
Column (1): Mask. (2): Minimum red intensity contrast threshold. (3): Maximum
CaIIK intensity contrast threshold.  (4): Corresponding magnetic flux density \citep[]{ras03,ort05}. (5): Percentage of pixels
retained by the mask.}
\end{deluxetable}

\clearpage

\clearpage

\clearpage

\clearpage

\clearpage

\clearpage

\clearpage

\clearpage

\end{document}